\documentstyle[11pt,newpasp,twoside,epsf]{article}
\def\edcomment#1{\iffalse\marginpar{\raggedright\sl#1\/}\else\relax\fi}
\reversemarginpar

\begin{document}
\title{The 6dF Galaxy Survey}

\author{Ken-ichi Wakamatsu}
\affil{Faculty of Engineering, Gifu University, Gifu 501-1192, Japan}
\author{Matthew Colless}
\affil{Mount Stromlo \& Siding Spring Observatories, ACT 2611, Australia}
\author{Tom Jarrett}
\affil{IPAC, Caltech, MS 100-22, Pasadena, CA 91125, USA}
\author{Quentin Parker}
\affil{Macquarie University, Sydney 2109, Australia}
\author{William Saunders}
\affil{Anglo-Australian Observatory, Epping NSW 2121, Australia}
\author{Fred Watson}
\affil{Anglo-Australian Observatory, Coonabarabran NSW 2357, Australia}

\begin{abstract}
The 6dF Galaxy Survey (6dFGS) \footnote{Member of Science Advisory Group: M. Colless 
(Chair; ANU, Australia), J. Huchra (CfA, USA), 
T. Jarrett (IPAC, USA), O. Lahav (Cambridge, UK), J. Lucey (Dahram, UK), G. Mamon 
(IAP, France), Q. Parker 
(Maquarie Univ. Australia),  D. Proust (Meudon, France), E. Sadler (Univ. Sydney, Australia), 
W. Saunders (AAO, Australia),  K. Wakamatsu (Gifu Univ., Japan), F. Watson (AAO, Australia)} 
is a spectroscopic survey of the entire southern
sky with $|b|>10\deg$, based on the 2MASS near infrared galaxy catalog. It is
conducted with the 6dF multi-fiber spectrograph attached to the 1.2-m UK 
Schmidt Telescope. The survey will produce redshifts for some 170,000 galaxies, and 
peculiar velocities for about 15,000 and is expected to be complete by June 2005. 
\end{abstract}

\section{Introduction}
In order to reveal large-scale structures at intermediate and large distances, extensive galaxy 
redshift surveys have been carried out, e.g. the 2dFGRS and SDSS, and the Hubble- and Subaru-deep 
field surveys.  There is now an urgent need to study the large-scale structure of the 
Local Universe that can be compared with the above deeper surveys. However to do this
required hemispheric sky coverage which can only be effectively and efficiently carried out
with a dedicated Schmidt telescope with a wide field of view. To this end the 
AAO implemented a 6dF multi-fiber spectrograph (Fig. 1) for the UK Schmidt Telescope (UKST)
and has now commenced a full southern hemisphere 
galaxy redshift survey of the Local Universe. In this paper an outline of the survey is briefly 
reported. Further details are given at the following web site:

http://www.mso.anu.edu.au/6dFGS/

\section{6dF: A Multi-Fiber Spectrograph on the UKST}
6dF is a multi-fiber spectrograph attached to the UKST.  
It is named after the telescope's field of view which is 6 degrees in diameter, 
just as 2dF the equivalent 2 degree mulit-fibre system at the 3.9m Anglo-Australian Telescope.
6dF consists of an automated fiber positioner and a fast $F/0.9$ CCD spectrograph.  
6dF is the third generation of multi-fiber spectrograph on the UKST 
and its early history is described in the Appendix.

\begin{figure}
\plotfiddle{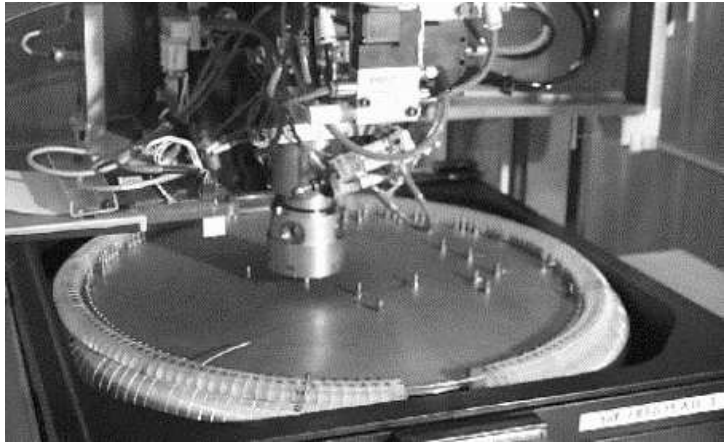}{5.5cm}{0}{60}{60}{-130}{-0}
\caption{6dF, an automated fiber positioner, configures magnetic fiber buttons on the curved 
focus of the field assembly under precise robotic control (5 $\mu$m ) at the exact
co-ordinates of celestial objects.}
\end{figure}

Each field assembly has 150 fibers of 100 $\mu$m core diameter, which corresponds to 6.7 arcsec on the 
sky. The 6dF positioner places magnetic fiber-buttons on the curved field plates (mandrel) which 
matches the telescope's curved focal surface (Fig. 1). The positioner operates off-telescope unlike 2dF.  
It takes less than 1 hour to accurately place 150 fibers including the defibering process from the
previous configuration. There are currently two field plate assemblies, so that 
one can be configured while the other is on the telescope. 
Further details of the system are given by Watson et al. (2001).

The 10m optical-fiber cable feeds the existing floor-mounted spectrograph, which has a 
Marconi 1024 $\times$ 1024 CCD detector with 13 $\mu$m pixels. Each spectrum is recorded on 
3 lines of CCD pixels.  The thinned CCD is back illuminated and has broad band coating for
enhanced blue sensitivity which is as high as 75 \% even at 3900 \AA, so redshifted H \& K lines 
are detected easily (Fig. 2).

\section{Performance of Survey Telescopes}
Performance of a survey telescope is simply expressed by how large a volume a telescope can cover in a 
given observing time. The sky area surveyed in a given exposure is represented by the telescope's 
field of view $\Omega$ in square degree, while an efficiency of the telescope
is given by a light-collecting area $A$, which is
proportional to square of an aperture of the telescope.  Hence, the survey performance of a telescope 
can be expressed by the $A\Omega$ product with the larger values indicating increasing survey power.  
These values for some typical telescopes are given in Table 1.

\begin{table}
\label{perform}
\caption{Performance of Survey Telescope}
\begin{tabular}{lccc} \tableline
{Telescope}&{Aperture}&{Field of View}&{$A\Omega$} \\ 
           &{(m)}       &{(degree)}  &{(m$^2\cdot$degree$^2$)}\\ \tableline 
Gemini  & 8.1 & 0.17 & 1.8 \\
WHT      & 4.2 & 0.5  & 4.4 \\
VLT      & 8.1 & 0.4  & 10  \\
Subaru  & 8.2 & 0.5 & 17  \\
Kiso Schmidt & 1.05 & 5.0 & 28 \\
Sloan   & 2.4 & 2.5 & 36  \\
UKST    & 1.2 & 6.0  & 52  \\
AAT     & 3.9  & 2.0  & 61  \\
 \tableline \tableline
\end{tabular}
\end{table}

Table 1 shows that the UKST has a very high survey performance due principally to its wide field 
$6 \deg$ diameter field of view, while the new generation of large aperture telescopes such
as Subaru and the VLT have relatively low survey performance though they can obviously
penetrate much deeper. Hence the UKST is ideal for wide-field but shallow surveys.

\section{6dF Galaxy Survey Design}

The 6dFGS was designed according to the following strategies:
\\
\quad {\bf Differentiation:}  What does the 6dFGS offer that is not offered by the \\
\qquad \quad 2dFGRS, SDSS, or other
surveys?
\\
\quad {\bf Impact:}  What survey characteristics are required in order to maximize \\
\qquad \quad the science impact?
 \\
\quad {\bf Timeliness:} How quickly must the survey be carried out in order
to achieve \\
\qquad \quad its goals in a timely and competitive manner?
 
\vspace{1mm}
The 6dFGS has two distinct components: {\it a redshift survey} and {\it a peculiar velocity survey}.  
Target selection is based not on optical galaxy photometric selection like the 2dFGRS and SDSS, 
but on $K$-band selection from the recently completed near infrared 
2MASS all sky survey (Jarrett et el. 2000b). Our survey area is 
for the entire southern sky with $|b|>10\deg$ and amounts to 17,000 deg$^{2}$.  For the redshift 
survey, the surface density of targets must match or exceed the density of 6dF fibres to allow 
efficient observing.  Allowing 10 fibers for sky, this means the sample should have a mean surface
density of at least 5 deg$^{-2}$. Furthermore target galaxies should be bright enough to allow 
redshifts to be measured in relatively 
short integration times so that the whole southern sky can be covered in a reasonable 
amount of time. We set the limiting magnitude at 12.75 in $K_{s}$-band\footnote{We adopted 
a corrected total magnitude $K_{tot}$ estimated from the isophotal magnitude $K_{20}$ given in the
2MASS catalog.}, and so finally selected about 120,000 objects from the 2MASS Extended Source Catalog.

\begin{figure}
\plotfiddle{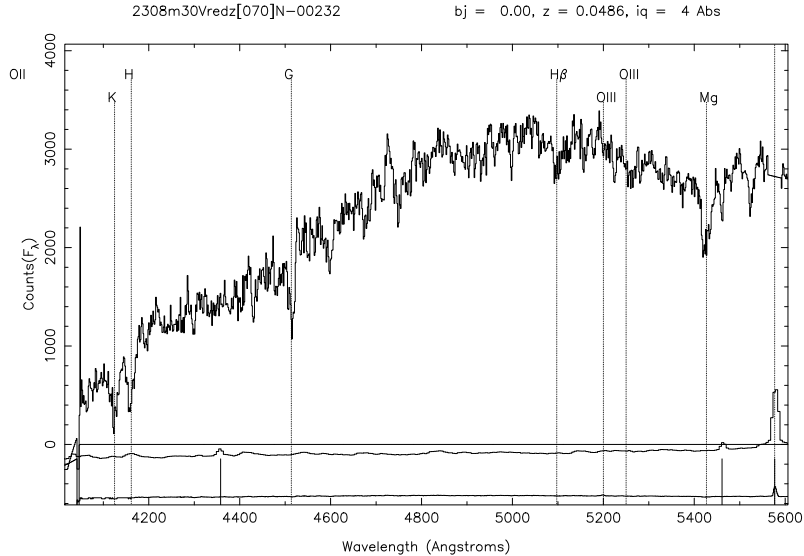}{7cm}{-90}{40}{40}{-150}{210}
\caption{An example of a blue spectrum of a galaxy of $z=0.0486.$ The H and K lines are 
clearly seen.}
\end{figure}

Observations for the survey are made with two different gratings for each target field (Table 2).
These
parameters are likely to change slightly with the imminent commissioning of volume-phase holographic gratings
in a new transmissive arrangement which offers enhanced system efficiency.

\begin{table}
\caption{Parameters of Spectroscopic Observations}
\begin{tabular}{lccccc} \tableline
{Spectrum}&{Grating}&{Spectral Coverage}&{FWHM}& {Exposure} \\ \tableline 
Blue  & 600V & 4000\AA \ - 5600\AA & 5 \ - 6 \AA & 3 $\times$ 20 min  \\
Red   & 316R & 5400\AA \ - 8400\AA & 9 \ - 12 \AA & 3 $\times$ 10 min \\ \tableline \tableline
\end{tabular}
\end{table}

Total exposure times for each survey field is about 1.5 hours.
Observing overheads
between fields 
takes a further
30 - 40 minutes. This permits 5 fields per night in the long winter lunations reducing to 3 in summer.

To observe target galaxies as efficiently as possible, 6dF field centers have been carefully 
determined from an adaptive tiling algorithm (Campbell, Saunders, \& Colless 2002). The total number of field centers 
is 1360.  These sky configurations cover 95\% of all the target galaxies, with 
an efficiency of 87\% usage of fibers.  In crowded regions 
like cluster centers, multiple observations with different fiber configurations are required to 
observe close pairs of galaxies, because fibers cannot be put within a minimum separation 5 arcmin 
due to the physical proximity constraints imposed by the footprint of the cylindrical magnetic buttons.

Altogether, the survey will produce redshifts for some 170,000 galaxies including objects for several additional target programs (see section 5.4).  
In addition, peculiar velocities will also be obtained for about 15,000 nearby bright early type Galaxies.  Both samples  will be complete by June 2005.  The first data release of the redshift survey is expected by the end of
2002.

\begin{figure}
\plotfiddle{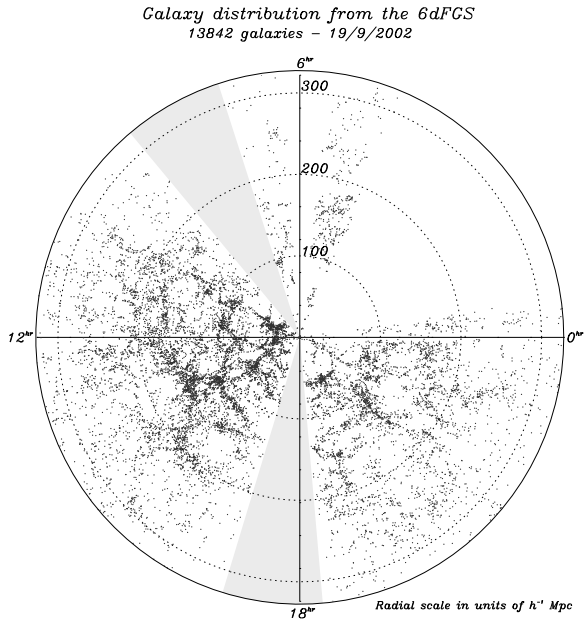}{8cm}{0}{50}{50}{-150}{-30}
\caption{Galaxy distribution from the 6dFGS for a strip $\delta =-30\deg$.}
\end{figure}

\section{Uniqueness of the 6dFGS}

The 6dFGS has the following characteristics as compared with 
2dFGRS, SDSS, and other redshift surveys.

\begin{enumerate}
\item {\em Near-Infrared Selection: A Clear Window on the Mass Distribution}
\\
Our survey is based on the new 2MASS near-infrared imaging survey of the whole sky. We 
use magnitude-limited 2MASS $J, H, \& K_{tot}$ galaxy samples, supplemented by complete photometric 
samples from other optical (SuperCOSMOS B \& R) and 
near-infrared (DENIS) catalogues. Near-infrared(NIR) luminosity, especially in the 
$K_s$-band, is not biased by recent star formation activity, and represents the stellar 
masses of individual galaxies more accurately than optical magnitudes which can be biased by star
formation activity. 
\\
\qquad Using NIR luminosity also minimizes the effects of internal absorption of galaxies, so that $M/L$ ratios 
are not affected by orientation, especially for spiral galaxies. 
\\
\qquad Absorption by dust in our own Galaxy is also much reduced so allowing 
greater sky coverage and more uniform sample selection over the whole 
sky (apart from a narrow region around the plane of the Milky Way). 
\\
\qquad Near-infrared target selection means that the 6dF Galaxy Survey is a very effective 
means of determining the true mass distribution in the Local Universe.

\item {\em Peculiar Velocities: Bulk Motions of Galaxies} \\ 
As well as measuring redshifts, we also measure galaxies' motions (their `peculiar
velocities'). We concentrate on early-type galaxies, and measure their peculiar 
velocities using the well-established $D_n$-sigma relation. 6dF allows us to obtain 
medium-dispersion, high-S/N spectra from which we can measure the central velocity 
dispersions of individual galaxies (Fig. 2). By comparing these dispersions with the galaxies' 
apparent sizes we can determine their distances. Combined with high quality redshift 
measurements, we can obtain peculiar velocities (the deviations from the Hubble flow) 
of significant numbers of individual galaxies.

\item {\em All-Sky Coverage: A Picture of the Local Universe} \\ 
The 2dF galaxy survey penetrates deep into space, but is limited to quite a 
narrow sky region around the south Galactic pole and a section of the celestial
equatorial zone, covering 5\% of the sky. The Sloan Digital Sky Survey is conducted 
with a dedicated 2.5m telescope, but covers less than one-third of the northern sky. 
The 6dF Galaxy Survey however covers the entire southern sky with galactic latitude greater 
than 10 degrees. This wide survey area is quite unique among the various current 
surveys and provides a full hemispheric description of the Local Universe.

\item {\em Additional Targets: Wide Windows to the Universe} \\
In the 6dFGS for fields having insufficient targets to use all 150 available fibers, remaining fibers have
been allocated to additional target programs according to scientific merit (such as
sampled from the ROSAT All-Sky Survey and NVSS radio survey). These various additional
target lists are combined in priority order with the main 6dFGS survey list to provide significant
added value to the original survey science
\end{enumerate}

\section{Scientific Aims}

The 6dFGS will provide a unique snapshot of the Local Universe based 
on a homogeneous, high-quality database (Fig. 3).   These data can be used in a wide range 
of scientific analyses.  \\
\quad The main scientific goals are:
\\
\qquad 1. The large-scale structure (density field) of the Local Universe. \\
\qquad 2. The bulk motions (velocity field) of galaxies in the Local Universe. \\
\qquad 3. The estimation of fundamental cosmological parameters, such as the \\
\qquad \quad mean mass density and cosmological constant, from the joint analysis of \\
\qquad \quad the density and velocity fields. \\
\qquad 4. The dependence of the properties of normal galaxies on their local envi- \\
\qquad \quad ronment and the surrounding large-scale structure. \\ 
\qquad 5. Studies of the properties of rare types of galaxies from additional target \\
\qquad \quad samples selected on the basis of their radio, far-infrared, optical or X-ray \\
\qquad \quad properties. 

\section{Survey in Zone of Avoidance}
At galactic latitudes below $|b| <10 $\deg, several important clusters 
and structures have been discovered, such as the Great Attractor and
the Ophiuchus cluster, one of the brightest X-ray cluster in the sky (Wakamatsu et al. 2000). 
The 2MASS Extended Source Catalog provides target galaxies even in this area despite high extinction and high
density of foreground stars (Jarrett et al. 2000a) so we can use 6dF to 
penetrate as deep as possible into the galactic plane over some limited area to
further study the extent and form of these and related important features.

\section{Future of Schmidt Telescopes in Spectroscopic Mode}
The 6dF spectroscopic survey mode has opened a 
new era for the UKST. 
As with all sky imaging surveys, it is also important to extend the 6dF spectroscopic survey 
into the northern sky. The Kiso Schmidt telescope is one of the best telescopes that could
accomplish this if equiped with a 6dF type system. 
There are already plans for a new innovation at the UKST that will allow more than 2000 fibers 
to be placed simultaneously on star positions to study the dynamics and chemical evolution of 
our Galaxy.

The implementation of a much bigger Schmidt telescope for performing muilti-fiber spectroscopy
(such as LAMOST in China) is a further extension of this trend.

\section*{Appendix: Astronomy with a Glue  -- Early History of 6dF}
There is a long history leading up to the implementation of the powerful new 6dF multi-fiber
spectrograph.  In the early 1980s when nobody imagined using a Schmidt telescope for spectroscopy,
Fred Watson started to put fibers on a curved focal plane with a precision of 20-50 $\mu$m. 


After many trials, Dr. Watson and his collaborators succeeded in putting fibers 
in the following manner (Fig. 4): \ i) a honey-comb mandrel 
plate holder was manufactured to have room for putting fibers from the backside of the 
mandrel, ii) a glass plate of a positive contact copy of a sky survey plate 
is set on the mandrel to use as a template for galaxy positions, iii) inserting 
fibers into a cut-down syringe needle and housing, and iv) attaching the fiber ferrule on 
the backside of glass plate with glue.
 The first spectrograph was made from a Pentax camera 
with hypersensitised Tech-Pan film, and was called FLAIR.  System throughput was very poor 
due to scattering of light on the template. 

\begin{figure}
\plotfiddle{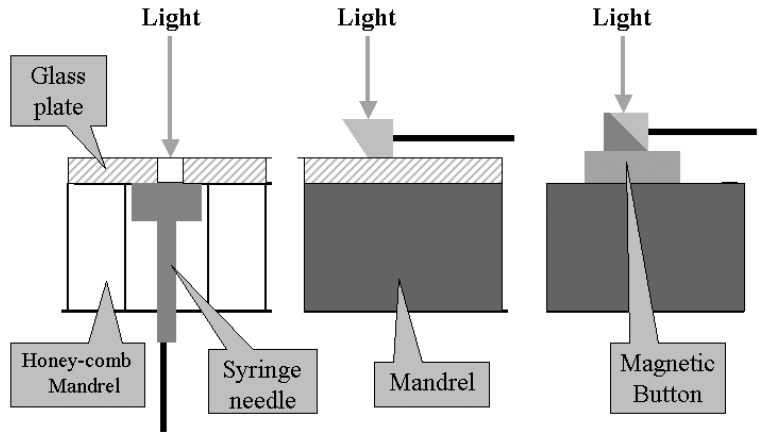}{5.cm}{0}{40}{40}{-140}{-0}
\caption{Schematic illustrations of set-up of an optical fiber on the field plate assembly 
for the original system (left), FLAIR II (middle), and 6dF (right), respectively.}
\end{figure}

FLAIR II, the second generation machine, was fabricated under a quite different design: 
i) a glass plate of a {\em negative} contact copy was set on a Mandrel, ii) fibers were 
connect via
a modified syringe needle mount or ferrule to a small right-angle prism, 
iii) with a semi-automated fiber 
positioner, the prism-ferrule assembly glued at a galaxy position on the glass plate using
UV curing cement with a precision of about 20 $\mu$m, and iv) 
92 fibers running complicatedly on the surface 
of glass plates were taped and bundled to prevent from blocking of light on the fiber.  
At this stage, a new spectrograph was fabricated with a fast optics and a CCD detector.  

FLAIR II yielded decent throughput, and yielded good performance allowng many useful 
science projects to be undertaken.  
However, it had a serious problem; it took 6-7 hours to put 92 fibers to 
target objects.  
To overcome this problem, the present day 6dF fully-automated fiber positioner 
robot was commissioned in June 2001 by Will Saunders and Quentin Parker.  

\acknowledgements
We thank all the members of Science Advisory Group.  We deeply express our 
thanks to Mr. M. Hartley and other stuff members at the AAO for their nice 
observations.  KW is supported by a grant-in-aid of Ministry of 
Education, Culture, Science \& Technology of Japan under a No. 13640236.


\begin{references}
\reference Campbell, L., Saunders, W., \& Colless, M.M. 2002,
in preparation
\reference Jarrett, T.H., et al. 2000a, \aj, 119, 2498
\reference Jarrett, T.H., et al. 2000b, \aj, 120, 298
\reference Wakamatsu, K., et al. 2000, in ASP Conf. Ser. Vol. 218, 
Mapping the Hidden Universe: The Universe Behind the Milky Way, 
eds.  R.C. Kraan-Korteweg, P.A. Henning, \& H. Andernach (San Francisco: ASP), 187
\reference Watson F.G., et al. 2001, in ASP Conf. Ser. Vol. 232, The New Era in Wide-Field Astronomy', 
eds, R. Clowes, A. Adamson, \& G. Bromage, (San Francisco: ASP), 421
\end{references}
\end{document}